\documentclass[prl,preprint,superscriptaddress,showpacs,amssymb,amsmath,amsfonts,aps]{revtex4}
\usepackage{graphicx}
\usepackage{epsfig}
\usepackage{dcolumn}

\def\be{\begin{equation}}
\def\ee{\end{equation}}
\def\bea{\begin{eqnarray}}
\def\eea{\end{eqnarray}}

\begin{document}

\title{Transverse momentum dependence of semi-inclusive pion production}
   
\newcommand*{\YERPHY}{Yerevan Physics Institute, Yerevan, Armenia}
\newcommand*{\JLAB}{Thomas Jefferson National 
Accelerator Facility, Newport News, Virginia 23606}
\newcommand*{\UMASS}{University of Massachusetts Amherst, 
Amherst, Massachusetts 01003}
\newcommand*{\RPI}{Rensselaer Polytechnic Institute, 
Troy, New York 12180}
\newcommand*{\NCSU}{North Carolina A \& T State University, 
Greensboro, North Carolina 27411}
\newcommand*{\BUCH}{Bucharest University, Bucharest, Romania }
\newcommand*{\ARGONNE}{Physics Division, Argonne National 
Laboratory, Argonne, Illinois 60439 }
\newcommand*{\HAMPTON}{Hampton University, Hampton, Virginia 23668 }
\newcommand*{\MARYLAND}{University of Maryland, College Park, 
Maryland 20742 }
\newcommand*{\GWU}{The George Washington University, 
Washington, D.C. 20052 }
\newcommand*{\VASS}{Vassar College, Poughkeepsie, 
New York 12604 }
\newcommand*{\VALJE}{Vrije Universiteit, 1081 HV Amsterdam, T
he Netherlands }
\newcommand*{\FIU}{Florida International University, 
University Park, Florida 33199 }
\newcommand*{\WITS}{University of the Witwatersrand, 
Johannesburg, South Africa }
\newcommand*{\HOU}{University of Houston, Houston, TX 77204 }
\newcommand*{\UVA}{University of Virginia, Charlottesville, 
Virginia 22901 }
\newcommand*{\MSS}{Mississippi State University, Mississippi 
State, Mississippi 39762 }
\newcommand*{\DUKE}{Triangle Universities Nuclear Laboratory 
and Duke University, 
Durham, North Carolina 27708 }
\newcommand*{\CALTECH}{California Institute of Technology, 
Pasadena, California 91125 }
\newcommand*{\UNC}{University of North Carolina Wilmington, 
Wilmington, North Carolina 28403 }
\newcommand*{\REGINA}{University of Regina, Regina, Saskatchewan, 
Canada, S4S 0A2 }
\newcommand*{\RUTGERS}{Rutgers, The State University of New Jersey, 
Piscataway, New Jersey, 08855 }
\newcommand*{\UCONN}{University of Connecticut, Storrs, 
Connecticut 06269 }
\newcommand*{\OHIO}{Ohio University, Athens, Ohio 45071 }
\newcommand*{\JMU}{James Madison University, Harrisonburg, 
Virginia 22807 }
\newcommand*{\GETTY}{Gettysburg College, Gettysburg, 
Pennsylvania 18103}

\author{H.~Mkrtchyan}  
\affiliation{\YERPHY}
\author{P.E.~Bosted} 
\affiliation{\JLAB}
\affiliation{\UMASS}
\author{G.S.~Adams} 
\affiliation{\RPI}
\author{A.~Ahmidouch}
\affiliation{\NCSU}
\author{T.~Angelescu}
\affiliation{\BUCH}
\author{J.~Arrington} 
\affiliation{\ARGONNE}
\author{R.~Asaturyan} 
\affiliation{\YERPHY}
\author{O.K.~Baker}
\affiliation{\JLAB}
\affiliation{\HAMPTON}
\author{N.~Benmouna}
\affiliation{\GWU}
\author{C.~Bertoncini}
\affiliation{\VASS}
\author{H.P.~Blok}
\affiliation{\VALJE}
\author{W.U.~Boeglin} 
\affiliation{\FIU}
\author{H.~Breuer}
\affiliation{\MARYLAND}
\author{M.E.~Christy} 
\affiliation{\HAMPTON}
\author{S.H.~Connell} 
\affiliation{\WITS}
\author{Y.~Cui}
\affiliation{\HOU}
\author{M.M.~Dalton} 
\affiliation{\WITS}
\author{S.~Danagoulian}
\affiliation{\NCSU}
\author{D.~Day}
\affiliation{\UVA}
\author{T.~Dodario} 
\affiliation{\HOU}
\author{J.A.~Dunne} 
\affiliation{\MSS}
\author{D.~Dutta}
\affiliation{\DUKE}
\author{N.~El~Khayari} 
\affiliation{\HOU}
\author{R.~Ent} 
\affiliation{\JLAB}
\author{H.C.~Fenker}
\affiliation{\JLAB}
\author{V.V.~Frolov} 
\affiliation{\CALTECH}
\author{L.~Gan} 
\affiliation{\UNC}
\author{D.~Gaskell}
\affiliation{\JLAB}
\author{K.~Hafidi} 
\affiliation{\ARGONNE}
\author{W.~Hinton} 
\affiliation{\HAMPTON}
\author{R.J.~Holt}
\affiliation{\ARGONNE}
\author{T.~Horn} 
\affiliation{\JLAB}
\author{G.~M.~Huber} 
\affiliation{\REGINA}
\author{E.~Hungerford}
\affiliation{\HOU}
\author{X.~Jiang} 
\affiliation{\RUTGERS}
\author{M.~Jones} 
\affiliation{\JLAB}
\author{K.~Joo}
\affiliation{\UCONN}
\author{N.~Kalantarians} 
\affiliation{\HOU}
\author{J.J.~Kelly} 
\affiliation{\MARYLAND}
\author{C.E.~Keppel} 
\affiliation{\JLAB}
\affiliation{\HAMPTON}
\author{V.~Kubarovsky} 
\affiliation{\JLAB}
\author{Y.~Li}
\affiliation{\HOU}
\author{Y.~Liang} 
\affiliation{\OHIO}
\author{S.~Malace}
\affiliation{\BUCH}
\author{P.~Markowitz} 
\affiliation{\FIU}
\author{E.~McGrath} 
\affiliation{\JMU}
\author{P.~McKee}
\affiliation{\UVA}
\author{D.G.~Meekins} 
\affiliation{\JLAB}
\author{B.~Moziak} 
\affiliation{\RPI}
\author{T.~Navasardyan}
\affiliation{\YERPHY}
\author{G.~Niculescu} 
\affiliation{\JMU}
\author{I.~Niculescu} 
\affiliation{\JMU}
\author{A.K.~Opper}
\affiliation{\OHIO}
\author{T.~Ostapenko} 
\affiliation{\GETTY}
\author{P.E.~Reimer} 
\affiliation{\ARGONNE}
\author{J.~Reinhold}
\affiliation{\FIU}
\author{J.~Roche} 
\affiliation{\JLAB}
\author{S.E.~Rock} 
\affiliation{\UMASS}
\author{E.~Schulte}
\affiliation{\ARGONNE}
\author{E.~Segbefia} 
\affiliation{\HAMPTON }
\author{C.~Smith} 
\affiliation{\UVA}
\author{G.R.~Smith}
\affiliation{\JLAB}
\author{P.~Stoler} 
\affiliation{\RPI}
\author{V.~Tadevosyan} 
\affiliation{\YERPHY}
\author{L.~Tang}
\affiliation{\JLAB}
\affiliation{\HAMPTON}
\author{M.~Ungaro} 
\affiliation{\RPI}
\author{A.~Uzzle} 
\affiliation{\HAMPTON}
\author{S.~Vidakovic}
\affiliation{\REGINA}
\author{A.~Villano} 
\affiliation{\RPI}
\author{W.F.~Vulcan} 
\affiliation{\JLAB}
\author{M.~Wang}
\affiliation{\UMASS}
\author{G.~Warren} 
\affiliation{\JLAB}
\author{F.~Wesselmann} 
\affiliation{\UVA}
\author{B.~Wojtsekhowski}
\affiliation{\JLAB}
\author{S.A.~Wood} 
\affiliation{\JLAB}
\author{C.~Xu} 
\affiliation{\REGINA}
\author{L.~Yuan} 
\affiliation{\HAMPTON}
\author{X.~Zheng} 
\affiliation{\ARGONNE}
\author{H.~Zhu} 
\affiliation{\UVA}

\newpage
\date{\today}


\begin{abstract}

Cross sections for semi-inclusive electroproduction of 
charged pions ($\pi^{\pm}$) from both proton and deuteron targets were
measured for $0.2<x<0.5$, $2<Q^2<4$ GeV$^2$, 
$0.3<z<1$, and $P_t^2<0.2$ GeV$^2$. 
For $P_t<0.1$ GeV, we find the azimuthal dependence to 
be small, as expected theoretically.
For both $\pi^+$ and $\pi^-$, the $P_t$ dependence from the 
deuteron is found to be slightly weaker than from the proton.
In the context of a simple model, this implies that the initial transverse
momenta width of $d$ quarks is larger than for $u$ quarks and, contrary to
expectations, the transverse  momentum width of the favored fragmentation
function is larger than the unfavored one.
 
\end{abstract}

\maketitle



    A central question in the understanding of nucleon structure is the 
orbital motion of partons. Much is known about the light-cone 
momentum fraction, $x$, and virtuality scale, $Q^2$, dependence 
of the up and down quark parton distribution functions (PDFs) 
in the nucleon. In contrast,  very little is 
presently known about the dependence of these 
functions on their transverse momentum $k_t$. 
Simply based on the size of the 
nucleon in which the quarks are confined, one 
would expect characteristic transverse momenta 
of order a few hundred MeV, 
with larger values at small Bjorken $x$ where 
the sea quarks dominate, and 
smaller values at high $x$ where all of 
the quark momentum is longitudinal in the limit $x=1$.  
Increasingly  precise studies of the nucleon 
spin sum rule~\cite{EMC,E155,HERMES,RHIC} strongly suggest 
that the net spin carried by quarks and gluons is relatively small, and 
therefore the net orbital angular momentum must be significant. This in
turn implies significant transverse momentum of quarks.
Questions that naturally arise include: 
what is the flavor and helicity dependence of the 
transverse motion of quarks and gluons, and 
can these be modeled theoretically
and measured experimentally?

\begin{figure}
\begin{center}
\epsfxsize=3.20in
\epsfysize=2.60in
\epsffile{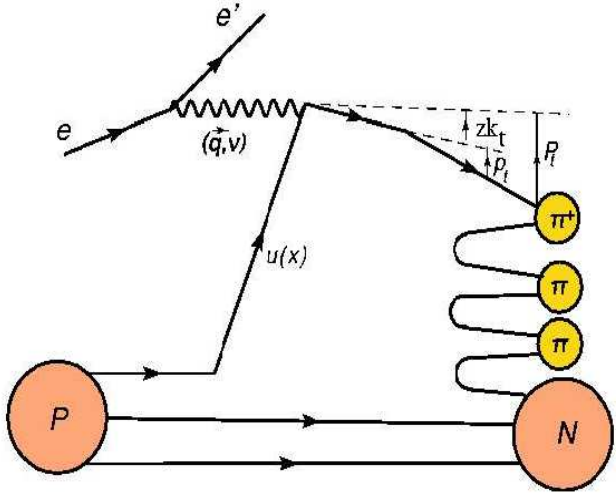}
\caption{\label{fig:sidis_diagram}
Schematic diagram of semi-inclusive pion 
electroproduction within a factorized
QCD parton model at lowest order in $\alpha_s$. Final transverse momenta of 
the detected pion $\vec P_t$ arises from convolving the struck quark 
transverse momenta $\vec k_t$ with the transverse momentum
generated during fragmentation process $\vec p_t$. }
\end{center}
\end{figure}

The process of semi-inclusive deep-inelastic lepton scattering (SIDIS),
$lN\rightarrow lhX$ has been shown to factorize~\cite{Ji04}, in the high 
energy limit, into lepton-quark scattering 
followed by quark hadronization. 
Ideally, one could directly measure the quark transverse momentum
dependence of the quark distribution 
functions $q(x,k_t)$ by detecting all 
particles produced in the hadronization process. 
In the present experiment, we detect only a 
single hadronization product: a 
charged pion carrying an energy fraction $z$ of the available energy.
The probability of producing a pion with a 
transverse momentum $P_t$ relative 
to the virtual photon ($\vec{q}$) direction 
is described by a convolution of the quark distribution functions  and 
$p_t$-dependent 
fragmentation functions $D^+(z,p_t)$ and $D^-(z,p_t)$,
where $p_t$ is the transverse momentum of the pion relative
to the quark direction, with the imposed 
condition~\cite{Anselmino}
$\vec{P}_t = z\vec{k}_t + \vec{p}_t$ (see 
Fig.~\ref{fig:sidis_diagram}).
The ``favored'' and ``unfavored'' functions  $D^+(z,p_t)$ 
and $D^-(z,p_t)$ refer to the case where the produced 
pion contains of the same flavor as 
the struck quark or not. 
``Soft'' non-perturbative processes are 
expected~\cite{Anselmino} to generate relatively small
values of $p_t$ with an 
approximately Gaussian distributions in $p_t$. Hard QCD processes
are expected to generate large non-Gaussian tails for $p_t>1$ GeV,
and probably do not play a major role in the interpretation of the
present experiment, for which the total transverse momentum
$P_t<0.45$ GeV. The assumption that the fragmentation functions
do not depend on quark flavor (for example $D^+(z,p_t)$ applies
equally well to $u\rightarrow \pi^+$ and $d\rightarrow \pi^-$) in
principle allows the $k_t$ widths of up and down quarks to 
be distinguished. In the present experiment, the use of both
proton and deuteron targets (the latter with a higher $d$ quark
content than the former) and the detection of both $\pi^+$ and
$\pi^-$ permits a first study of this problem.

\begin{figure}
\begin{center}
\epsfxsize=3.20in
\epsfysize=2.60in
\epsffile{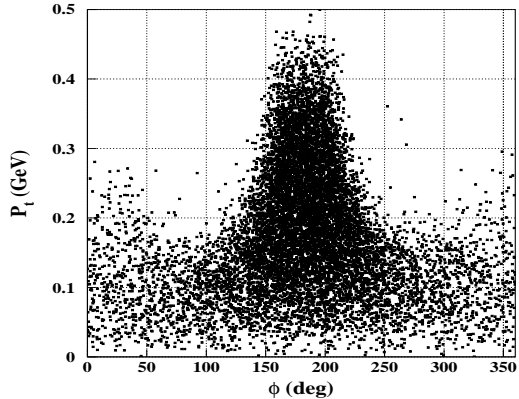}
\caption{\label{fig:phipt}
$P_t$ distribution of data from this experiment as a function
of $\phi$.}
\end{center}
\end{figure}

  The experiment (E00-108) used the Short Orbit (SOS) and High 
Momentum (HMS) spectrometers in Hall C at 
Jefferson Lab to detect final state electrons and pions, respectively.  
An electron beam with energy of 5.5 
GeV and currents ranging between 20 and 60 
$\mu A$ was provided by the CEBAF accelerator.
Incident electrons were scattered from 4-cm-long liquid hydrogen or
deuterium targets.  The experiment consisted 
of three parts: i) at a fixed 
electron kinematics of ($x,Q^2$) = (0.32, 2.30 GeV$^2$), 
$z$ was varied from 0.3 to 1, with nearly 
uniform coverage in the pion azimuthal 
angle, $\phi$, around the virtual photon 
direction, but at a small average $P_t$ 
of 0.05 GeV; ii) for $z=0.55$, $x$ was varied from 0.2 to 0.5 (with a 
corresponding variation in $Q^2$, from 2 to 4 GeV$^2$), 
keeping the pion 
centered on the virtual photon direction (and again average $P_t$ 
of 0.05 GeV);  
iii) for ($x,Q^2$) = (0.32, 2.30 GeV$^2$), $z$ 
near  0.55, $P_t$ was scanned 
from 0 to 0.4 GeV by increasing 
the HMS angle (with average $\phi$ near 180 
degrees). The $\phi$ distribution as a function of $P_t$
is shown for all three data sets combined in 
Fig.~\ref{fig:phipt}. The virtual photon-nucleon invariant mass $W$, 
was  always larger 
than 2.1 GeV (typically 2.4 GeV), corresponding to the traditional 
deep inelastic region for inclusive scattering.

At lower virtual photon energy and/or 
mass scales, the factorization ansatz is expected to 
break down, due to the effects of final state interactions, 
resonant nucleon
excitations, and higher twist contributions~\cite{Melnit01}. 
In particular, in the present experiment the residual invariant mass 
$M_x$ of the undetected particles 
(see Fig.~\ref{fig:sidis_diagram}) ranges
from about 1 to 2 GeV (inversely correlated with $z$), spanning the mass 
region traditionally associated with significant 
baryon resonance excitation. 
The extent to which this situation leads to a break-down of 
factorization was 
studied in our previous paper~\cite{Nav07}. 
It was found that good agreement with expectations based on 
higher energy data was achieved for $z<0.7$, approximately 
corresponding to $M_x > 1.5$ GeV. 
The ratio of total up to down quark distributions 
$u(x)/d(x)$ extracted from 
ratios of cross sections, as well as the ratio of valence-only up to down 
ratios  $u_v(x)/d_v(x)$, were also found to be reasonably 
compatible with higher energy extractions, provided $z<0.7$. 
This issue will be addressed further for the $P_t$-scan data below.
Finally, the ratio of unfavored 
to favored fragmentation functions 
$D^-(z)/D^+(z)$ (from the $\pi^-/\pi^+$ 
ratios on the deuteron) was found to be consistent with extractions 
from other experiments. All of these studies 
were done with the $z$-scan and 
$x$-scan data, for which the average $P_t$ was small ($<0.1$ GeV ), 
and the average value of $\cos(\phi)$ was close to zero.


  In this paper, we focus on the $P_t$ dependence, with the 
goal of searching 
for a possible flavor dependence to the quark 
distribution functions and/or 
fragmentation functions. Since the average value 
of $\cos(\phi)$ in the present experiment is 
correlated with $P_t$ (approaching -1 for the largest $P_t$ value of 
0.45 GeV, see Fig.~\ref{fig:phipt}), we first study the limited 
data available from this experiment on
the $\phi$ dependence, which must be an even 
function since neither the beam
nor the target were polarized. We 
parameterize~\cite{Dakin73} the data 
for each target and pion flavor according to:
\begin{equation}
\label{eq:factorization}
{d\sigma_{ee'\pi x} \over d\sigma_{ee'x}}=
\frac{dN}{dz}\, b\, \exp (-bP^{2}_t)\, \frac{1+A\cos \phi +
B\cos (2\phi )}{2\pi }\, 
\end{equation}
where the parameters $A(x,Q^2,z,P_t)$ and 
$B(x,Q^2,z,P_t)$ are a measure of 
the relative importance of the interference 
terms ${\sigma}_{LT}$ and 
${\sigma}_{TT}$, respectively~\cite{donnelly}. 
The assumed Gaussian $P_t^2$ dependence
(with slopes $b$ for each case) is an effective parameterization that
seems to describe the data adequately for use in making radiative
and bin-centering corrections. We use this model for studying the $\phi$
dependence, then return to a more detailed study of the
$P_t$ dependence in the context of a simple model that incorporates
a different  $P_t$ dependence for each struck quark and produced
hadron flavor. 

\begin{figure}
\begin{center}
\epsfxsize=4.00in
\epsfysize=4.00in
\epsffile{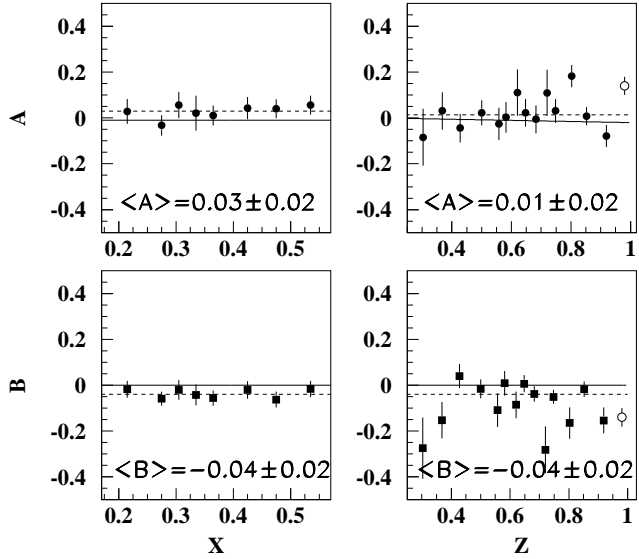}
\caption{\label{fig:ab_xz_scan}
The parameters $A$ and $B$ [the relative coefficients of 
the $\cos{\phi}$ ($\sigma_{LT}$) and $\cos{2\phi}$ 
($\sigma_{TT}$) terms] averaged over $\pi^+$ and 
$\pi^-$ detected from proton and deuteron targets,
as a function  of $x$ at $\langle z\rangle $=0.55 (left), and 
as a function of $z$ at $\langle x\rangle $=0.32 (right). 
The average value of transverse momentum
($\langle \vert P_t\vert \rangle $) is $\sim$0.05 GeV. 
The dashed lines indicate the weighted averages for $z<0.7$, which 
are also enumerated in each panel. Errors indicated include only
statistical contributions. Systematic errors are highly
correlated from point to point, and are estimated at 0.03
on both $A$ and $B$. The open symbols are from 
exclusive pion production (see text). The solid lines are
theoretical predictions~\protect\cite{Cahn}.}
\end{center}
\end{figure}

For each kinematic point in the $x$ and $z$ scans
(average $P_t=0.05$ GeV, maximum $P_t$ 0.2 GeV), 
we extracted $A$  and $B$ and found no
statistically significant difference between the 
results for $\pi^+$ or 
$\pi^-$, or proton or deuteron targets. We therefore 
combined all four cases 
together, and present the results in Fig.~\ref{fig:ab_xz_scan}. 
Systematic errors (not shown in the figure) are approximately
0.03 on both $A$ and $B$ and are highly correlated from
point to point. Taking the systematic errors into account, 
 the values of $A$ and $B$ are close to  zero, for all 
values of  $x$ studied, and for values of $z<0.7$, 
where our previous studies 
showed a good consistency with factorization.
The small values of $A$ and $B$ are also 
consistent with the expectations based on kinematic shifts due to parton 
motion as described by Cahn~\cite{Cahn} (shown 
as the solid curves on the 
figures) and Levelt-Mulders~\cite{Levelt-Mulders}.
These effects are proportional to $P_t$ for $A$, and $P^2_t$ for $B$
respectively~\cite{Cahn,Levelt-Mulders,Berger,Oganes}, 
so are suppressed at low $P_t$. More specifically, 
using the assumption that
the average quark and fragmentation widths are equal, the Cahn~\cite{Cahn}
asymmetries are given by 
\begin{eqnarray}
\label{eq:AB}
A=-\gamma (2<P_t>/Q)(2-y)\sqrt{1-y}/[1+(1-y)^2],\\
B=-\gamma^2(2<P_t^2>/Q^2)(1-y)/[1+(1-y)^2],
\end{eqnarray}
where $\gamma=z^2/(1+z^2)$, $y=\nu/E$, $\nu$ is the virtual
photon energy, and $E$ is the beam energy, yielding
$A=-0.01$ and $B=-0.0002$ for $z=0.55$.
The more recent treatment of Ref.~\cite{Anselmino} also gives
results for $A$ and $B$ which are very close to zero
(especially for $B$).
Other possible higher twist contributions will also be 
proportional to powers 
of $P_t/\sqrt(Q^2)$~\cite{Metz,Bacchetta}, and therefore
suppressed at our lower average values of $P_t$ and $P_t^2$.
Specifically, the twist-2 Boer-Mulders~\cite{Boercalc}
contribution to $B$ is essentially zero 
in the models of Ref.~\cite{Boercalc,Vincenzo}.

  In contrast, the  longitudinal-transverse and transverse-transverse 
coefficients $A$ and $B$ are much larger in exclusive pion production
($M_x = M$, where $M$ is the nucleon mass) than those predicted  for 
SIDIS. This is evidenced by our extracted 
average values for exclusive 
$\pi^\pm$ electroproduction on deuteron and for $\pi^+$ on 
proton, shown as the open symbols near ${z=0.98}$ in 
Fig.~\ref{fig:ab_xz_scan}.
This underlines the importance of accounting for the radiative tail from 
exclusive production, which in our analysis was done using the 
computer code EXCLURAD~\cite{Exclurad} together with a 
reasonable model of exclusive pion electroproduction.
The corrections where checked with the Hall C simulation
package SIMC, which treats radiative corrections
in the energy and angle peaking approximation~\cite{SIMC}.


\begin{figure}
\begin{center}
\epsfxsize=4.00in
\epsfysize=4.00in
\epsffile{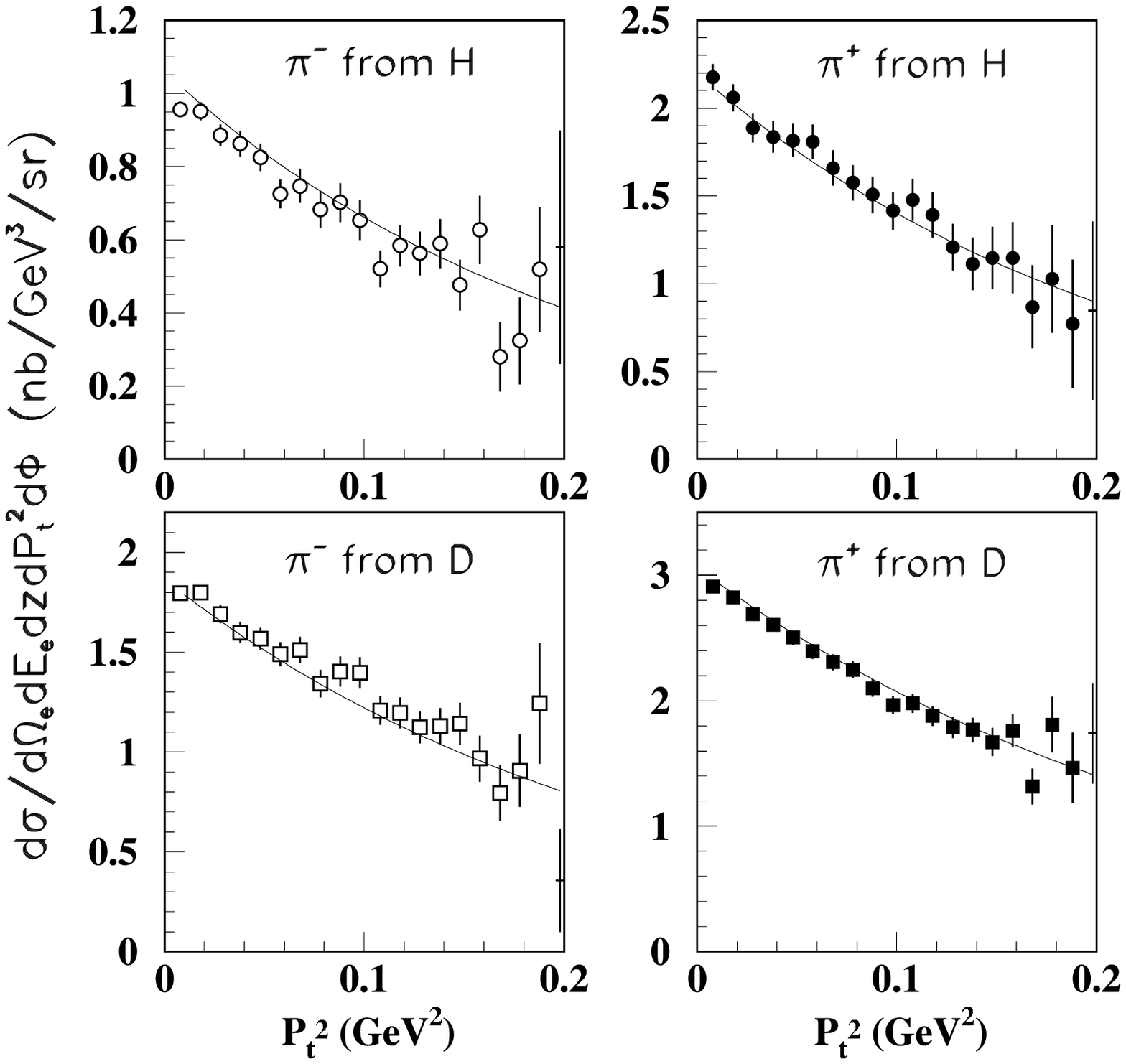}
\vspace{-0.2in}
\caption{\label{fig:slope_pt} The $P^2_t$ dependence of 
differential cross-sections per nucleus 
for $\pi^\pm$ production on 
hydrogen (H) and deuterium (D) targets at $\langle z\rangle $=0.55 and $\langle x\rangle $=0.32. The solid lines show
the result of the seven-parameter fit described in the text.
 The error bars are statistical only. Systematic errors are
typically 4\% (relative, see text for details).
The average value of $\cos(\phi)$ varies with $P_t^2$
(see Table~1.}
\end{center}
\end{figure}

   We now turn to the study of the $P_t$ scan data. 
We used the cross section model from our previous 
paper~\cite{Nav07} to describe the $Q^2$ dependence of the data
(needed because $P_t$ and $Q^2$ are somewhat correlated),
and extracted cross sections at fixed $Q^2$ averaged over $\phi$.
The corrections for  $Q^2$ dependence did not distinguish
between targets or pion flavor.
Relatively small corrections (typically a few percent) 
for radiative effects (including
the tails from exclusive pion production) and diffractive 
$\rho$ production were made~\cite{Nav07} for each
case individually. The systematic error on these corrections
is estimated to be approximately 2\%. The normalization errors
due to target thickness, computer and electronic dead time,
beam charge measurement, beam energy, and spectrometer
kinematics combine to approximately 2\% overall, and 1\%
from case to case. The overall error due to spectrometer
acceptance is estimated to be 3\%, but $<1\%$ from case
to case because targets were exchanged frequently, as was 
the spectrometer polarity. 
The extracted cross sections are 
shown in Fig.~\ref{fig:slope_pt}
and listed in Table~\ref{tab:xsects_vrs_pt2}. The acceptance-averaged
values of $\cos(\phi)$ range from -0.3 at low $P_t$ to nearly -1
at higher $P_t$, while the average 
values of $\cos(2\phi)$ approaches
1 at high $P_t$ (See Fig.~\ref{fig:phipt} and   
Table~\ref{tab:xsects_vrs_pt2}).


\begin{figure}
\begin{center}
\epsfxsize=4.0in
\epsfysize=3.0in
\epsffile{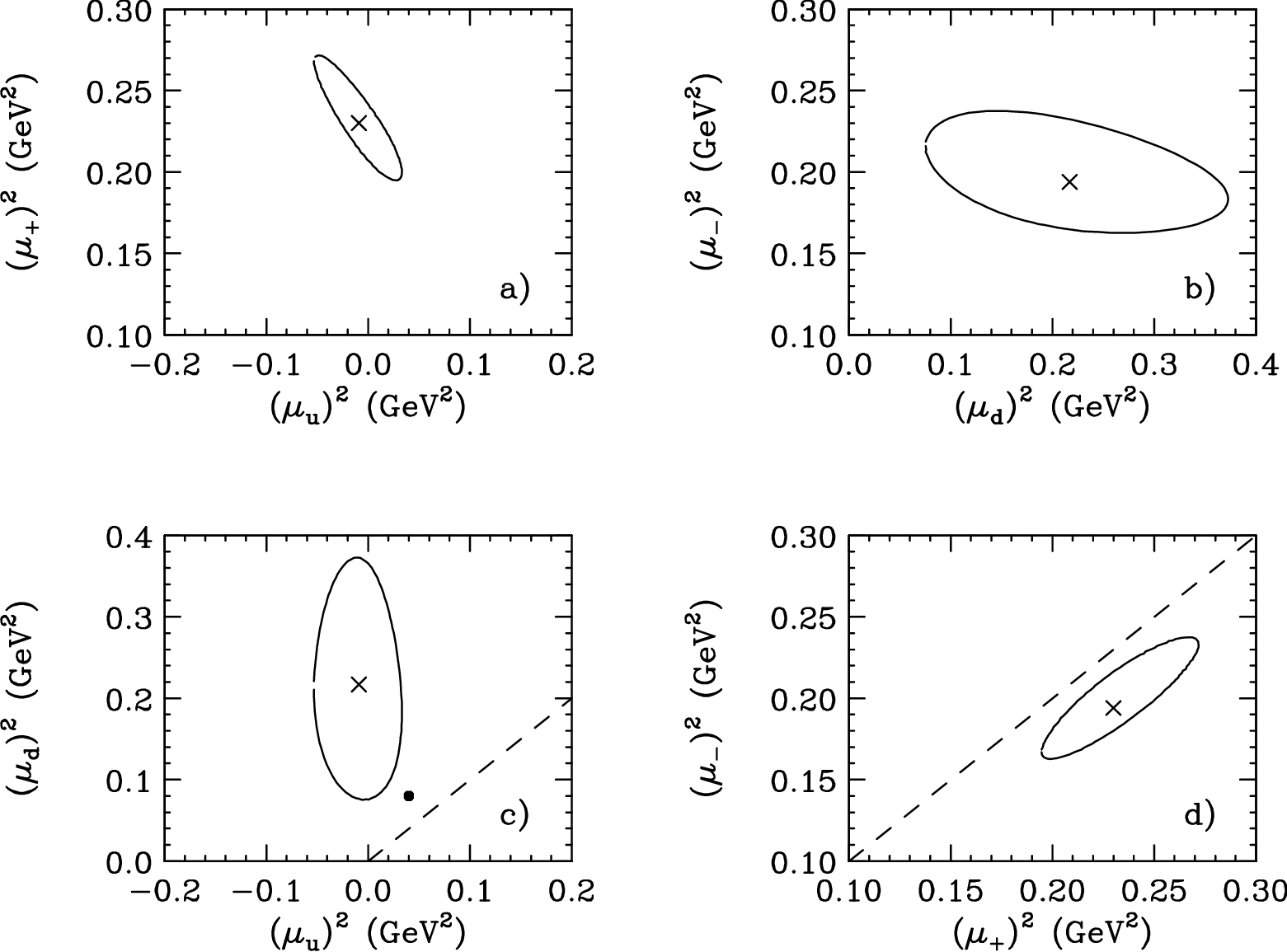}
\caption{\label{fig:fit}
Fit parameters (crosses) and two-standard-deviation 
contours from the seven-parameter 
fit to the data shown in 
Fig.~3: a) $u$ quark width squared $\mu_u^2$ versus favored 
fragmentation width squared $\mu_+^2$; 
b) $\mu_d^2$ versus $\mu_-^2$; 
c) $\mu_u^2$ versus $\mu_d^2$; 
d) $\mu_-^2$ vs $\mu_+^2$. 
The large dot near the bottom of panel c  
is from a di-quark model~\protect\cite{Jakob}. The dashed line
in panels c and d indicate $\mu_u^2=\mu_d^2$ and 
$\mu_-^2=\mu_+^2$, respectively.}
\end{center}
\end{figure}

Examination of Fig.~\ref{fig:slope_pt} shows that the 
$P_t$-dependence for $\pi^+$ and $\pi^-$ are
very similar to each other for each target, 
but that the slopes for
the deuteron target are somewhat  smaller
than those for the proton. 
For a more quantitative understanding
of the possible implications, we study the data in the context of
a simple model in which the $P_t$ dependence is described in
terms of two Gaussian distributions for each case. 
Following Ref.~\cite{Anselmino}, we assume
that the widths of quark and fragmentation functions are Gaussian
in $k_t$ and $p_t$, respectively, and that the convolution of
these distributions combines quadratically. The main difference
from Ref.~\cite{Anselmino} is that we allow separate widths
for up and down quarks, and separate widths for favored and
unfavored fragmentation functions. The widths of 
the up and down distributions are given by $\mu_u$ and $\mu_d$,
respectively, and the favored (unfavored) fragmentation widths
are given by $\mu_+$ ($\mu_-$). Following Cahn~\cite{Cahn} and
more recent studies~\cite{Anselmino}, we assume that only the
fraction $z$ of the quark transverse momentum contributes to 
the pion transverse momentum (see Fig.~\ref{fig:sidis_diagram}).
 We assume further 
that sea quarks are negligible (typical 
global fits show less than 10\% contributions at $x=0.3$).
To make the problem tractable, we take only the leading
order terms in $(k_t/Q)$, which was shown to be a reasonable
approximation for small to moderate $P_t$ in Ref.~\cite{Anselmino}.
The simple model can then be written as:
\begin{eqnarray}
\label{eq:sidis_xsects}
\left.
\begin{array}{lll}
\sigma_p^{\pi +} = C   [4 c_1(P_t) e^{-b_u^+ P_t^2} + 
                          (d/u) (D^-/D^+) c_2(P_t) e^{-b_d^- P_t^2}] \\
\sigma_p^{\pi -} = C   [4  (D^-/D^+) c_3(P_t) e^{-b_u^- P_t^2} + 
                          (d/u) c_4(P_t)e^{-b_d^+ P_t^2}] \\
\sigma_n^{\pi +} = C   [4 (d/u) c_4(P_t)e^{-b_d^+ P_t^2} + 
                           (D^-/D^+) c_3(P_t)e^{-b_u^- P_t^2}] \\
\sigma_n^{\pi -} = C   [4 (d/u) (D^-/D^+) c_2(P_t)e^{-b_d^- P_t^2} + 
                           c_1(P_t)e^{-b_u^+ P_t^2}] \\
\end{array}
\right.
\end{eqnarray}
\noindent where $C$ is an arbitrary normalization factor, and 
the inverse of the total widths for each combination of quark
flavor and fragmentation function are given by
\begin{eqnarray}
\label{eq:b}
\left.
\begin{array}{lll}
b_u^\pm=(z^2\mu_u^2 + \mu_\pm^2)^{-1} \\
b_d^\pm=(z^2\mu_d^2 + \mu_\pm^2)^{-1} \\
\end{array}
\right.
\end{eqnarray}
and we assume
$\sigma_d = (\sigma_p + \sigma_n$)/2. 
The Cahn effect~\cite{Cahn,Anselmino}  is taken into 
account through the terms:
\begin{eqnarray}
\label{eq:c}
\left.
\begin{array}{lll}
c_1(P_t) = 1. + c_0(P_t,<\cos(\phi)>) \mu_u^2 b_u^+ \\
c_2(P_t) = 1. + c_0(P_t,<\cos(\phi)>) \mu_d^2 b_d^- \\
c_3(P_t) = 1. + c_0(P_t,<\cos(\phi)>) \mu_u^2 b_u^- \\
c_4(P_t) = 1. + c_0(P_t,<\cos(\phi)>) \mu_d^2 b_d^+ \\
c_0(P_t,<\cos(\phi)>) = \frac { 4 z (2-y) \sqrt{1-y}}
 { \sqrt{Q^2} [1 + (1-y)^2 ] } \sqrt{P_t^2} <\cos(\phi)>. \\ 
\end{array}
\right.
\end{eqnarray}
We fit for the four widths ($\mu_u$, $\mu_d$, $\mu_+$, and
$\mu_-$), $C$,
and the ratios $D^-/D^+$ and $d/u$, where the fragmentation
ratio is understood to represent the data-averaged value at $z=0.55$,
and the quark distribution ratio is understood to represent
the average value at $x=0.3$. The fit describes the data
reasonably well ($\chi^2=78$ for 73 degrees of freedom), and
finds the somewhat low ratio  $d/u=0.30\pm 0.03$ 
(the LO GRV98 fit~\cite{GRV} has about 0.40 for valence
quarks), and the more reasonable ratio
$D^-/D^+=0.42\pm 0.01$ (a fit 
to HERMES results~\cite{Geiger}, 
$D^-/D^+=1/(1+z)^2$, predicts 0.42 at $z=0.55$). 
Both $d/u$ and $D^-/D^+$ are 
largely uncorrelated with other fit parameters. Since the
data are at fixed $z$, the main contributions that distinguishes
large fragmentation widths from large quark widths are the
$\phi$-dependence Cahn-effect $c_i$ terms. While  there is
a significant inverse correlation between the two most 
important quark and fragmentation widths, 
($\mu_u$ and $\mu_+$, respectively), the fit finds a clear
preference for $\mu_u$ to be smaller 
than $\mu_+$ as shown in Fig.~\ref{fig:fit}a. On the other
hand, the fit finds $\mu_d$ and $\mu_-$ to be of the same
magnitude and not strongly correlated, as shown
in Fig.~\ref{fig:fit}b.

The fit tends to favor a larger $k_t$ width
for $d$ quarks ($\mu_d^2=0.22\pm 0.13$ GeV$^2$) than for 
$u$ quarks ($\mu_u^2=-0.01\pm 0.04$ GeV$^2$), as illustrated in 
Fig.~\ref{fig:fit}c, although the error on the $d$ quark width
is relatively large. The tendency is 
consistent with a di-quark model~\cite{Jakob} 
in which the $d$ quarks are only found in an axial 
di-quark, while the $u$ quarks are predominantly
found in a scalar di-quark. If the axial and scalar 
di-quarks have different masses, for example 0.9 and 
0.6 GeV, then the $d$ quark
distribution falls off more slowly with $k_t$ than the $u$ quark
distribution. In this model, the distributions show considerable
deviation from an exponential falloff, but if we take the 
slope at the origin, the corresponding widths are 
 $\mu_u^2=0.04$ and $\mu_d^2=0.08$ GeV$^2$, 
as illustrated by the solid dot in Fig.~\ref{fig:fit}c. 
Fixing the quark widths to these values still gives a reasonable
fit to our data ($\chi^2=81$ for 75 degrees of freedom).
The magnitude of both widths 
is moderately sensitive to the choice of the model parameter
$\lambda_0$ (we used 0.6 GeV), although the difference in widths
is largely driven by the difference in axial and scalar 
di-quark masses (for example, increasing the axial di-quark 
mass to 1.4 GeV increases  $\mu_d^2$ to 0.25 GeV$^2$, 
which is the central value of our fit). Using the fit
parameters, we find the magnitude of the
$\cos(\phi)$ term $A$ (Eq.~\ref{eq:AB}) at $P_t=0.4$ GeV to
be about -0.05 for all cases except $\pi^+$ production from
hydrogen, where it is 0.01. These results are similar in
sign and magnitude to those found in the HERMES
experiment~\cite{HermesPhi}.

We find that the fragmentation widths $\mu_+$ and $\mu_-$ are
correlated, as illustrated 
in Fig.~\ref{fig:fit}d, although the allowed range is not large,
and the central values ($\mu_+^2=0.23\pm 0.04$ GeV$^2$ and 
$\mu_-^2=0.19\pm 0.04$ GeV$^2$) are in reasonable agreement with
both each other and also the flavor-averaged
value of 0.20 GeV$^2$ found in Ref.~\cite{Anselmino}. 
While there is a slight
tendency for the favored width to be larger than the 
unfavored one, a reasonable fit can be obtained setting
the widths equal to each other ($\chi^2=81$ for 74 d.f.,
$\mu_+^2=\mu_-^2=0.22\pm0.04$ GeV$^2$).

To estimate the effect of experimental systematic errors on our fit
results, we repeated the fits with: no diffractive $\rho$
subtraction; no exclusive radiative tail subtraction; 
relative target thickness changed by 1\%; and difference 
in $\pi^+$ and $\pi^-$ absorption changed by 1\%. In all
cases, the quark and fragmentation width 
results remained well within the error ellipses
shown in Fig.~\ref{fig:fit}. The only parameter that changed
significantly is the $d/u$ ratio, which goes up to 0.33
with no exclusive tail subtraction. We found no significant
change to the fit parameters upon 
adding to $\mu_u^2$ and $\mu_d^2$ an average nucleon 
transverse momentum
squared of 0.001 GeV$^2$ (evaluated using  the Paris wave 
function~\cite{Paris}) for the deuteron model. 

  In summary, we have measured semi-inclusive electroproduction of 
charged pions ($\pi^{\pm}$) from both proton and deuteron targets, using 
5.5 GeV energy electrons at Jefferson Lab.
We find the azimuthal dependence to be small, compared to
exclusive pion electroproduction, and consistent with
theoretical expectations~\cite{Cahn, Anselmino}. 
In the context of a simple model with only valence quarks and only 
two fragmentation functions, we 
find the $k_t$ width of $d$ quarks
to be larger than for $u$ quarks, for which the width is
consistent with zero within the statistical error. 
We find that the favored 
fragmentation $p_t$ width to be (unexpectedly) somewhat larger than
the unfavored width, although both are
larger than the $u$ quark width.

All of the above fit 
results can only be considered as suggestive
at best, due to the limited kinematic range covered, the
somewhat low $u/d$ ratio that we find, and 
the very simple model assumptions described above.
Many of these limitations could be
 removed with future experiments covering
a wide range of $Q^2$ (to resolve additional higher twist
contributions), full coverage in $\phi$, a larger
range of $P_t$, a wide range in $z$ (to distinguish quark width terms, 
which is weighted by powers of $z$, 
from fragmentation widths, which likely vary
slowly with $z$), and including the $\pi^0$ final
state for an additional consistency check (particularly on the
assumption that only two fragmentation functions are needed
for charged pions from valence quarks). Some of these goals 
should be attainable with existing and upcoming data from
Jefferson Lab, especially after the planned energy upgrade.
These data should  provide potential
information on how hadron transverse momentum in SIDIS
is split between fragmentation and intrinsic quark
contributions. 

\medskip
The authors wish to thank H. Avakian, A. Afanasev, and M. Schlegel 
for useful discussions, and the Physics Letters referees for 
useful suggestions. This work is supported in part by research 
grants from the U.S. Department 
of Energy and the U.S. National Science Foundation.
The Southeastern Universities Research Association operates the
Thomas Jefferson National Accelerator Facility under the
U.S. Department of Energy contract DEAC05-84ER40150.


\begin{table}
\caption{\label{tab:xsects_vrs_pt2} Differential 
cross-sections per nucleus for $\pi^{\pm}$
production on hydrogen and deuterium versus $P^2_t$. 
The error bars are statistical only. The 
values of $\cos(\phi)$ and $\cos(2\phi)$ averaged
over the experimental acceptance are also indicated
(see Fig.~\protect{\ref{fig:phipt}}).
}
{\centering  \begin{tabular}{c c c c c c c}
$P^2_t$ & $<\cos(\phi)>$ & $<\cos(2\phi)>$ &  
$\sigma^{\pi^+}_p$ & $\sigma^{\pi^-}_p$& 
$\sigma^{\pi^+}_d$ & $\sigma^{\pi^-}_d$ \\
GeV$^2$  & & & nb/sr/GeV$^3$ & nb/sr/GeV$^3$ &
 nb/sr/GeV$^3$  & nb/sr/GeV$^3$ \\
\hline
  0.008 &  -0.369 &   0.031 &   2.177$\pm $ 0.075 &   0.956$\pm $ 0.021 &   2.912$\pm $ 0.038 &   1.796$\pm $ 0.030 \\
  0.018 &  -0.511 &   0.089 &   2.058$\pm $ 0.077 &   0.951$\pm $ 0.024 &   2.824$\pm $ 0.040 &   1.800$\pm $ 0.037 \\
  0.028 &  -0.533 &   0.105 &   1.885$\pm $ 0.082 &   0.885$\pm $ 0.030 &   2.688$\pm $ 0.045 &   1.690$\pm $ 0.045 \\
  0.038 &  -0.875 &   0.580 &   1.834$\pm $ 0.089 &   0.863$\pm $ 0.035 &   2.602$\pm $ 0.051 &   1.599$\pm $ 0.052 \\
  0.048 &  -0.892 &   0.623 &   1.815$\pm $ 0.094 &   0.825$\pm $ 0.038 &   2.504$\pm $ 0.055 &   1.567$\pm $ 0.056 \\
  0.058 &  -0.935 &   0.761 &   1.808$\pm $ 0.097 &   0.726$\pm $ 0.040 &   2.393$\pm $ 0.060 &   1.491$\pm $ 0.060 \\
  0.068 &  -0.941 &   0.780 &   1.658$\pm $ 0.100 &   0.747$\pm $ 0.047 &   2.307$\pm $ 0.063 &   1.511$\pm $ 0.067 \\
  0.078 &  -0.946 &   0.799 &   1.575$\pm $ 0.101 &   0.683$\pm $ 0.050 &   2.247$\pm $ 0.065 &   1.344$\pm $ 0.069 \\
  0.088 &  -0.952 &   0.818 &   1.507$\pm $ 0.105 &   0.702$\pm $ 0.053 &   2.099$\pm $ 0.069 &   1.403$\pm $ 0.074 \\
  0.098 &  -0.963 &   0.858 &   1.414$\pm $ 0.109 &   0.653$\pm $ 0.055 &   1.964$\pm $ 0.071 &   1.398$\pm $ 0.077 \\
  0.108 &  -0.963 &   0.860 &   1.477$\pm $ 0.120 &   0.520$\pm $ 0.050 &   1.980$\pm $ 0.075 &   1.208$\pm $ 0.073 \\
  0.118 &  -0.965 &   0.866 &   1.391$\pm $ 0.129 &   0.584$\pm $ 0.056 &   1.878$\pm $ 0.079 &   1.196$\pm $ 0.077 \\
  0.128 &  -0.963 &   0.857 &   1.208$\pm $ 0.133 &   0.563$\pm $ 0.060 &   1.789$\pm $ 0.085 &   1.123$\pm $ 0.080 \\
  0.138 &  -0.972 &   0.892 &   1.112$\pm $ 0.150 &   0.589$\pm $ 0.067 &   1.768$\pm $ 0.098 &   1.131$\pm $ 0.090 \\
  0.148 &  -0.972 &   0.892 &   1.146$\pm $ 0.176 &   0.476$\pm $ 0.069 &   1.671$\pm $ 0.111 &   1.142$\pm $ 0.105 \\
  0.158 &  -0.973 &   0.897 &   1.147$\pm $ 0.203 &   0.627$\pm $ 0.093 &   1.762$\pm $ 0.133 &   0.967$\pm $ 0.115 \\
  0.168 &  -0.975 &   0.902 &   0.868$\pm $ 0.236 &   0.280$\pm $ 0.095 &   1.316$\pm $ 0.145 &   0.795$\pm $ 0.139 \\
  0.178 &  -0.977 &   0.911 &   1.027$\pm $ 0.307 &   0.324$\pm $ 0.119 &   1.810$\pm $ 0.222 &   0.906$\pm $ 0.182 \\
  0.188 &  -0.977 &   0.911 &   0.771$\pm $ 0.366 &   0.519$\pm $ 0.171 &   1.465$\pm $ 0.280 &   1.244$\pm $ 0.303 \\
  0.198 &  -0.977 &   0.911 &   0.847$\pm $ 0.509 &   0.579$\pm $ 0.319 &   1.740$\pm $ 0.398 &   0.357$\pm $ 0.258 \\
\end{tabular}\par}
\end{table}

\end{document}